# Science 3.0: Corrections to the Science 2.0 paradigm

Vladimir B. Teif

BioQuant and German Cancer Research Center (DKFZ),
Im Neuenheimer Feld 267, 69120 Heidelberg, Germany
E-mail: vladimir.teif@bioquant.uni-heidelberg.de

## ABSTRACT

The concept of "Science 2.0" was introduced almost a decade ago to describe the new generation of online-based tools for researchers allowing easier data sharing, collaboration and publishing. Although technically sound, the concept still does not work as expected. Here we provide a systematic line of arguments to modify the concept of Science 2.0, making it more consistent with the spirit and traditions of science and Internet. Our first correction to the Science 2.0 paradigm concerns the "open access" publication models charging fees to the authors. As discussed elsewhere, we show that the monopoly of such publishing models increases biases and inequalities in the representation of scientific ideas based on the author's income. Our second correction concerns post-publication comments online, which are all essentially non-anonymous in the current Science 2.0 paradigm. We conclude that scientific post-publication discussions require special anonymization systems. We further analyze the reasons of the failure of the current post-publication peer-review models and suggest what needs to be changed in Science 3.0 to convert Internet into a large "journal club".

## INTRODUCTION

The term "Science 2.0" was introduced around 2008 to describe online-based medium for research, documentation and collaboration in analogy with the "Web 2.0" term coined for the description of the next generation of internet. At that time, several influential journals such as *Science*, *Nature* and *Scientific American* endorsed the use of this term and encouraged scientists to move online (1-4) and internet domains such as science20.com started appearing. One of the main features of Science 2.0 is the global networking facilitated by the internet. This feature can be already seen: many science bloggers from the US and Europe are already connected in one network, the think tank of the future science-online community. At the present time there are several thousands of Science 2.0 bloggers. This number can be estimated from the amount of scientists subscribed to online networking groups devoted to Science 2.0 at web sites such as LinkedIn.com (700 members), ResearchGate.net (~12,500 members). For example, Figure 1 shows the saturating dynamics of the number of subscribers of one of the first such online group "The Life Scientists" at Friendfeed.com. This limited number and tight connectedness allowed a lot of coordination in writing about "Science 2.0".

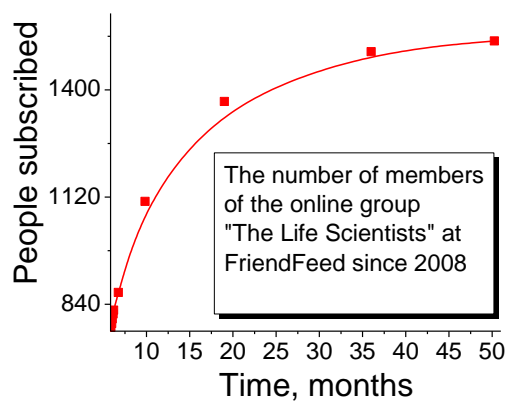

Figure 1. The number of subscribers of the group "The Life Scientists" at FriendFeed, as of 11.01.2013.

Currently accepted views about Science 2.0 can be summarized as follows (5):

1) Online networking is good because it multiplies efforts of many people and adds complementary expertise.

2) Online data sharing is good because it facilitates the process of discovery and is a more effective way of spending taxpayer's money.

3) Open-access publishing is good because it provides free access to professional articles for everyone. Open-access will be the only publishing model in future.

4) Online sharing of unfinished works, unpolished thoughts and critic is good because it allows any scientist to expose his/her opinion and receive credit for it. An honest, fearless researcher always putting his name under all his writings in the internet is the Science 2.0 hero.

The first two points are quite evident and, therefore, we will not discuss them in detail. Instead, we will concentrate on the last two points to show that their current understanding needs corrections. In addition, we distinguish another point of Science 2.0, which is apparent for those who closely watches online processes, although it has not been clearly articulated yet:

5) The virtual "Republic of Science", connecting worldwide researchers online, has been *de facto* created, and it operates by the rules of direct democracy rather than the rules of any individual governing body, something that the inventor of this term Michael Polanyi could not foresee in 1962 (6). This will become important in our discussions since the conceptual features of Science 3.0 are generic, applicable to any country and any scientific field.

**1. Science 3.0 still needs peer-review.**
Since peer-review is at the core of the functioning of the current scientific system, a lot of people have been thinking about ways to improve it. One radical view that only few people endorse is that peer-review is not needed at all (7). This perspective comes with the idea of self-publishing, either on a personal web site/blog, or using public repositories such as [ArXiv.org](ArXiv.org). [ArXiv.org](ArXiv.org) successfully functions for several decades; it is common for physicists and mathematicians to upload there the drafts of their manuscripts before submitting to peer-reviewed journals. In addition, there are examples of extraordinary good works ending at online archives and not published in peer-reviewed journals at all. Perhaps the best known example of this kind is the work of [Grigory Perelman](Grigory Perelman), who solved a long-standing mathematical problem of great importance and published the solution online at [ArXiv.org](ArXiv.org) (8). He was recently awarded the Fields medal (the highest award in mathematical sciences, which he refused to accept). He never submitted this paper to a peer-reviewed journal.

Many Science 2.0 proponents go further and consider blogs as promising tools for self-publishing. Aggregated scientific blog systems have been created, including thousands of personal blog, such as [scientificblogging.com](scientificblogging.com), [blogs.nature.com](blogs.nature.com), [researchblogging.org](researchblogging.org), [scienceblog.com](scienceblog.com), and [scienceblogs.com](scienceblogs.com). One of the main problems with self-publishing is that the amount of information increases tremendously, and so does the "information noise". Within a narrow field, a good personal taste and connections might still help to find the balance between the must-read and may-read articles. However, a few steps out of the scientific niche the scientist finds himself surrounded by unknown names and myriads of potentially useful works that cannot be explored in the whole life (9).

Classical peer-review journals have many problems, but they are good at decreasing the level of information noise by preventing obvious nonsense and violations of the scientific ethics. Since life is short and Internet is addictive, before investing the valuable time into reading a proposed paper, scientists would usually check that they are familiar either with the name of the author or the name of the journal, or that they know the research institution where the work was done or that they at least know the publisher. As discussed elsewhere, peer-review has a potential danger to become the peer-censorship for a specific journal or a group of journals (10). Therefore, different types of alternative peer review systems have been proposed, such as the non-anonymous peer-review prior to publication (e.g. at [*Biology Direct*](Biology Direct) and [*Frontiers*](Frontiers)) or post-publication peer-review (11).

Whatever is the mechanism of pre-publication peer-review, it provide a time-saving tool (but not more than that -- as with the stock market, the value of the stock does not necessarily reflect the performance of the company). An additional parameter which is difficult to predict for all players is the pressure from emerging scientific countries and new journals which would significantly change the citation distribution (12).

To summarize this section, we see that peer-review will be retained in Science 3.0, but will feel strong pressure from self-publishing. In order to win this competition, publication models will undergo some changes, as detailed below.

## 2. The "open-access" publishing model leads to biases and inequalities in the idea selection.

The common consensus in Science 2.0 is that all journals should be accessible online. In addition, free access to journals is highly coveted. The idea of having articles available freely to readers is called "open-access". In a more narrow sense, open-access is also a business model which is based on charging the publications costs to the authors instead of the readers. This model has also been applied long ago in the advertisement industry. In fact, several journals have explicitly printed in the past that, "page charges for this article have been partially paid by the authors, and the publication should therefore be considered as an advertisement". Those days are now gone, and paying for your article to be published is considered as a rule rather than an exception in Science 2.0.

Open-access publications have become quite popular, partly because it is widely believed that open-access articles usually get more citations (13) (this statement has been recently questioned (14,15)), and partly because open-access is now encouraged at many levels. Essentially, the terms open-access and Science 2.0 are sometimes even interchanged. The problem is that while the open-access business model is looked upon favorably by readers, it also has its serious caveats for authors. In particular, the current open-access costs for one paper are comparable to the average monthly income of a person in US/Europe, and the situation is even worse for the majority of other countries. With regard to these large fees, only few countries have adopted funding systems where the author is compensated for both research and publishing. The list of these countries will hardly increase, because the countries which do not profit from their own high-impact journals have few reasons to bail out foreign publishers. Furthermore, even in the countries which have adopted such funding systems, not all scientists have access to them. It is frequently written in the journal rules that the journal would consider publishing an article for free if the author cannot pay. However, in practice the editors of open-access journals are under pressure to avoid free articles. Although journals have some limited funds to give waivers to authors who cannot pay, yet they are still a business, and their bottom line would suffer if this were to be their regular activity. Young scientists and scientists who have only modest budgets would typically avoid such journals, thus creating an income-based bias for scientific ideas, which is inacceptable.

What are the alternatives to the open-access business model, allowing everyone's free access to the articles? Many publishers grant free access to the papers published several months ago. Further, due to common several-year gaps between the discovery and its implementation in medicine or technology, a several-month delay would not actually make papers outdated for the lay audience not involved in the intense scientific competition. Another possibility is for governments to subsidize open-access journals making them completely free both for the reader and the author. Importantly, publishing in the journal should be free for *all* its authors. If the journal allows waivers only to *some* authors, an income-based bias mentioned above remains. Finally, another possibility to allow public access to scientific articles is through the system of public libraries, as it was in the USSR more than two decades ago. In this scenario, internet era libraries can provide citizens' online access from their homes. Countries that cannot afford public libraries can be granted free or low-fee access by the publishers. For example, the journal PNAS had granted free access to 139 low-income countries. PNAS had nothing to lose since these countries would not pay for subscriptions anyway, yet they produced a lot of articles citing PNAS.

## 3. Post-publication comments and discussions require online hubs and anonymization systems.

Post-publication peer-review (11) as well as readers' comments at web sites of online journals (16) have been proposed as essential constituents of Science 2.0. Recent publications in specialist journals increasingly argue in favor of the establishment of post-publication peer-review systems (17), while this idea is still opposed by the major high-impact journals (18). Most importantly, scientists still do not use even existing commenting systems available at many online journals. Why not? The general idea was quite simple. In the traditional science, journals accept "comments" on the articles, subject to the journal's approval, which is usually at the editor's discretion. This is a very time-consuming process, which requires the authors to prepare a well-written text, and then the editor decides on the acceptance. Finally, a technical editor needs to work on the page layout. It is a serious work and responsibility for everyone involved, not surprising that comment articles are quite infrequent in the traditional journals. Comments online should have dramatically facilitate this process. A scientist just reads the article online, clicks the "comment" button and adds a couple of lines, e.g. that equation 15 is incorrect, or the figure caption is misplaced, or there is some fundamental problem with the method, or a literature reference is missing. Usual internet forums receive from 1 to 10 comments per 100 reads of the article, this ratio being roughly constant for a given forum (statistics, collected by the author). Based on such statistics, one would predict at least several comments being provided for each of the scientific articles online, since they already have thousands of reads weeks after publication. However, this is not what happens in reality. Most articles have no comments at all, even those which are highly disputable. Why is it so?

Let us look at the internet discussions in general. What we know from non-scientific internet forums is that the most democratic and open discussions occur when people have the option to remain anonymous. These are the old internet traditions. Even in the recent internet history, the most authoritative online collaborative tool [Wikipedia](#) is anonymous, while attempts to create analogous non-anonymous common knowledge tools such as [Google Knol](#) have failed so far. On the other hand, non-anonymous online social networking sites provide a new twist in the internet history. Several projects such as [LinkedIn.com](#), [ResearchGate.net](#), [Academia.edu](#), [Nature Network](#), [Mendeley.com](#) tried to use social networks for scientific collaborations. It was shown that many fruitful discussions take place in the informal, relaxed atmosphere of closed groups in social networking sites. However, scientists become more reluctant when it comes to the exposure of their opinion to the "whole internet" under their real name instead of a nickname.

While there are known open-science projects where participants decided to open to the public completely, such as the [Polymath](#) project devoted to mathematics, the general tendency is that scientists are reluctant to exhibitionism. For example, a relatively old online group "[Genomics: Next Generation DNA Sequencing (NGS) and Microarray](#)" at LinkedIn, consisting of almost 7,000 professionals including a lot of senior scientists from both academy and industry, was discussing for about two weeks the new option offered by LinkedIn to open the group content to the public. Not all people agreed with the argument that it is safe to open up if their surnames and profiles will not be visible. Up to now the group remains closed.

To understand the basis for the cautiousness with respect to the real-name policy, let us forget for a minute about the internet and return to the traditional science. It appears that scientists actually used to comment anonymously in most cases when this requires criticism. Disclosing the real name is incompatible both with the anonymous peer-review system and the anonymous voting system (the basis of the current understanding of democracy). Not surprisingly, most scientific internet forums with intense discussions are anonymous.

Following are several examples of popular internet forums in the field of molecular biology: [biology-online.org](#), [protocol-online.org](#), [molecularstation.com](#), [biotechniques.com](#), [SEQanswers.com](#), and [molbiol.ru](#). Each of these forums has around 20,000 users, covering, in total, approximately 100,000 molecular biologists. It happens that some of the users on these forums know each other's identities, but in general all these forums are anonymous. Anonymity allows asking stupid questions and getting quick professional answers; exchange ideas without revealing your current or nearest-future plans; peer-to-peer sharing of published papers; and honest evaluations of the works of the others. Anonymity also presents some inherent problems. For example, we cannot rely on the authority of the scientist who provided the answer. But that is in line with the basic science functioning: The validity of the arguments should not depend on the name of their author. A recent analysis of college students' perceptions and interpretations of internet credibility confirms that this is currently the prevailing point of view (19).

Scientific forums are very different from the journals. An important lesson we learn from them is that most discussions in the internet happen either in closed groups or under nicknames. However, if we do not disclose our name, we are not getting recognition for our contribution, which is the driving force of science (20). This is the point where internet is very different from science. What forces internet users to spend their time making comments at professional web sites? If we ask a question on a forum, we may derive a benefit directly from the answer. If we answer or comment on someone's answer by adding more details, there is still a possibility to learn, especially if we expect that someone will comment after us, checking our arguments. In addition, many people comment because they have an emotional motivation to do so.

Consequently, are we motivated enough to comment on a scientific article? In analogy with forums, the answer is definitely "yes". A student can ask a question and get an answer directly from the authors (e.g. if they are getting automatic email alerts for each posted comment) or from someone else who happened to read the same paper and found the question interesting. That would be useful for the others who will come months or years later, and will see some Frequently Asked Questions already answered. For those who are at the same level of expertise with the authors, online comments are more an opportunity to express their opinion and check whether other scientists feel the same about specific details of the article. Almost any paper has some weaknesses or points difficult to understand, which can be resolved by comments. (This is also true for the current manuscript!) There is usually no reason commenting if we agree with everything or understand everything. We comment if we have something to say or to ask. In this case, we are professionally and emotionally motivated and do not need additional profit of identifying ourselves, like in the examples with internet forums.

Anonymity allows for minimal efforts for the commenter. Importantly, it allows checking that our own arguments are right (or wrong) in a risk-free way to gain something from the discussion and not to reveal even a slight incompetence. This means that low quality comments can also arise, which is normal. The quality of comments should be regulated by the [Netiquette](#) (internet ethics), not by the science ethics (21). Inappropriate comments violating the Netiquette can be simply reported and removed. More than 10-year experience of the author with scientific forums tells that there are actually not so many situations when moderation is required. Furthermore, existing non-anonymous comments e.g. at *Nature* are of reasonably high quality. A recent quantitative study of the statistics of non-anonymous comments made at *PLoS Journals* (22) and *BMC Journals* (23) further supports this point, reporting just around 1% of comments as spam.

Now let us look from the point of view of the authors of the article that is being discussed online. From the first glance there could be fears that one day someone can find our mistake and openly dismiss online our work, trashing our efforts, time and money, something that is much less likely to happen in the traditional science system. One possibility to relieve these fears is to allow the authors a full moderation control over the discussion thread devoted to their article. However, careful thinking shows that the benefits of open comments significantly outweigh potential risks. Indeed, in the worst case we risk losing mere months of work rather than years (in the case if no one would point out to our mistake early enough). Most importantly, we will have a prompt interactive feedback (we could also have it through personal contacts, but internet does this without filtering, faster and more efficiently). Finally, it is nice to have a chance to promote our article in a world-wide "journal club" of its readers with questions, answers, comments and interactive discussions that will be valuable many years after the publication.

Returning to the web sites of online journals that exist at the moment, we see that in most cases comments are not allowed at all. In the non-scientific internet there are analogies to this behavior. Comments are usually allowed under news articles, but prohibited under paid advertisement-type articles. This is understandable, since someone has paid for the advertisement and does not want comments to interfere with its message. However, scientific articles are not advertisements (at least, they are not supposed to be). Those few scientific journals that allow comments online take precautions: they force the user to register, mandatorily indicating his identity and institutional address. Technically, this takes some time. Every additional second spent on the web site decreases chances that a busy scientist will keep his intention to comment. More importantly, the mandatory user registration makes postings non-anonymous. (Many journals explicitly prohibit anonymous postings).

To address these issues, several networking sites performed attempts to establish post-publication discussions away from the publishers, such as the projects [Papercritic.com](#) and [Plasmyd.com](#). In addition, users of commercial bibliographic software such as [EndNote](#) and [Papers](#) have the option to share their reading lists and comments with other users of this program online. Unfortunately, these comments are not linked directly to the journal web sites and therefore they might be unnoticed by the majority of scientists who read articles online. Furthermore, due to the intrinsic non-anonymity of social networking, such systems have difficulties in solving the main issue pointed out above, namely that critical non-anonymous comments are not natural both for Science and Internet. Therefore, massive online commenting will have to wait until journals allow the option of anonymous comments without registration. Furthermore, the journals would probably need to implement a special system erasing the history of the commenter's IP addresses or at least ensuring that this information remains strictly confidential. In addition, a system allowing basic forum features would be needed to insert quotes, images and upload files. Most importantly, the missing culture of anonymous/pseudonymous comments online should be established. Of course, not all comments will be anonymous, since in many cases it make sense to put your real name. At least, this should be optional.

CONCLUSIONS

Internet evolves very quickly and so does science. Most of the features considered as revolutionary several years ago are now either trivial or have been tried and did not work out. For example, it is quite interesting to read today an article entitled "The future of medical journals…" written in 1998 (24). Not so long ago, but still before the digital journal era, the authors were able to foresee many of the features that we already observe today. Hopefully, the predictions made in the current manuscript will be also realized in the near future. We have concluded that the pre-publication peer-review will survive as the way to ensure the quality control check, but will be complemented by self-publishing at online preprint repositories and by the post-publication comments on the articles. We have shown that the open-access publication model leads to scientific biases based on the author's income. We have also provided a systematic argumentation showing that Science 3.0 components including post-publication discussions in the form of world-wide "journal clubs" at web sites of online journals will require special anonymization systems. These are obviously not the only new features of Science 3.0, and the futuristic analysis should continue.

Several recent publications have indicated with surprise that "the old systems in both research and publishing prove to be more resistant to change than many online evangelists originally had anticipated" (25). Many members of the Science 2.0 community believe that the majority of their fellow scientists are old-fashioned or not enough informed and that is the only reason for not using the Science 2.0 tools. From the analysis performed here it appears that the situation is quite different: existing Science 2.0 tools require significant conceptual changes to become really useful for scientists. In fact, scientists of all ages are traditionally among the most active users of modern technologies. An old person checking his email using a mobile phone on the street is very likely to be a usual university professor going for a lunch between the lectures. Scientists are ready and eager to take everything new that has proved to be useful for their work. If something is not taken up massively, it just has not proved to be useful yet, indicating that some changes are needed for Science 2.0 tools.

In addition to the mostly sociological aspects mentioned above, this analysis suggests several hints which could be useful from the point of view of the science policy. One of these hints is that funding bodies can try new systems of grants for the scientists and institutes who are interested in serving for non-profit open-access journals owned by the government, instead of supporting commercial third-party open-access publishers. This would allow a full control over the free access of both the authors and the readers, and could be even cheaper for the society (in the current open-access system public money are paid both as grants to open-access publishers, and as grants to the authors to pay open-access fees to the publishers). In fact, most journals currently listed in the open-access directory at [doaj.org](doaj.org) are not the journals utilizing the open-access business model, but rather the journals published by research institutes on the governmental money, free for the readers and authors, and completely free from the business component. There is also an important legal component in the issue with anonymous comments online. Scientific comments at online journals cannot be treated in the same way as the customers' feedback on commercial products. The customers' feedback is required by law to be non-anonymous in many countries; otherwise, the developer of the criticized product can sue the web site owner instead of the anonymous commenter. Such legal requirement cannot be imposed on scientific comments, which might require special legal amendments.

A reader familiar with the Science 2.0 concepts might be disappointed that the picture outlined above seemingly goes away from the current state of the art of Science 2.0, not making use of many fashionable Web 2.0 features which are all essentially non-anonymous. But it should be understood that Web 2.0 is also not something frozen. It would be overly naïve to think that the Facebook-type behavior is the top level of the evolution of Internet. The next era of Internet, the Web 3.0, will be very different from Web 2.0, most probably including sophisticated privacy-protecting systems. There could be different understanding of the privacy online. It is different for teenagers, for scientists, for politicians… The privacy of a scientist is the privacy of his thinking. For many scientists who work seriously on a problem, non-anonymous sharing of the current reading lists and comments with the whole internet is a breach of privacy comparable to putting a web-camera in a bathroom. It is worth to note that historically, science was leading Internet, not the other way round. Therefore Science 3.0 should take the same role for the conceptual defining of Web 3.0, requesting new internet features rather than adopting itself for the existing ones.

ACKNOWLEDGEMENTS

The "Science 3.0" term was proposed in April 2008 in discussions of Vladimir Teif with Daniel Teyf and Mike Dye. The idea of anonymous post-publication comments was introduced by the user *genereg* at FriendFeed's group The Life Scientists in 2009. This lively discussion with more than 100 comments is still available [online](online). Fabian Erdel, Tatjana Andrasiuk, Yevhen Vainstein, Soenke Bartling, Daniel Teyf, Cindy Lair, Jean-Louis Sikorav and Matthias Schäfer provided many helpful comments on the previous version of the manuscript (26). Marco A. Javarone helped performing online networking experiments at the closed forums at generegulation.info. Many participants of *Friendfeed, ResearchGate, Molbiol.ru, LinkedIn* and *SEQanswers* online communities helped shaping the arguments expressed above. Comments on the second draft of the manuscript will be acknowledged in the final version submitted for publication. Vladimir Teif is supported by the Center for Modelling and Simulation in the Biosciences (BIOMS) and the German Cancer Research Center (DKFZ). This particular publication was not paid by any grant.